\documentclass[a4paper]{article}

\usepackage[T1]{fontenc}
\usepackage[utf8]{inputenc}
\usepackage{ae,aecompl}
\usepackage{graphicx}
\usepackage{url}
\usepackage{todonotes}
\usepackage{hyperref}
\usepackage{booktabs}
\usepackage{listings}
\usepackage{amsmath}
\usepackage{amssymb}
\usepackage[acronym]{glossaries}
\usepackage[sort,compress]{cite}
\usepackage{xspace}
\usepackage{caption}
\usepackage{a4wide}
\usepackage{subcaption}
\usepackage{pgfplots}
\usepackage{paralist}
\usepackage{textcomp}
\usepackage[noend]{algpseudocode}
\usepackage{algorithm}
\usepackage{lipsum}
\usepackage{xcolor}
\usepackage{colortbl}
\usepackage{multirow}

\definecolor{darkcandyapplered}{rgb}{0.64, 0.0, 0.0}
\definecolor{darkcoral}{rgb}{0.8, 0.36, 0.27}
\definecolor{darkcyan}{rgb}{0.0, 0.55, 0.55}
\definecolor{darkgoldenrod}{rgb}{0.72, 0.53, 0.04}

\pgfplotsset{compat=newest}
\usepgfplotslibrary{groupplots}
\usetikzlibrary{patterns}

\newcommand{\ttt}[1]{\texttt{#1}}

\newcommand{\tsc}[1]{\textsc{#1}}
\newcommand{\ie}{i.e.\xspace}
\newcommand{\eg}{e.g.\xspace}

\newcommand{\sys}{\tsc{Thyme}\xspace}

\def\papertitle{I Know What You Did Last Summer: Time-Aware Publish/Subscribe for Networks of Mobile Devices}
\def\paperauthor{João A. Silva, Hervé Paulino, João M. Lourenço, João Leitão, Nuno Preguiça}

\hypersetup{
	pdftitle={\papertitle},
	pdfauthor={\paperauthor},
}

\title{\papertitle}

\author{
João A. Silva\\
\ttt{jaa.silva@campus.fct.unl.pt}
\and
Hervé Paulino\\
\ttt{herve.paulino@fct.unl.pt}
\and
João M. Lourenço\\
\ttt{joao.lourenco@fct.unl.pt}
\and
João Leitão\\
\ttt{jc.leitao@fct.unl.pt}
\and
Nuno Preguiça\\
\ttt{nuno.preguica@fct.unl.pt}\\
\\
NOVA Laboratory for Computer Science and Informatics,\\
Departamento de Informática,\\
Faculdade de Ciências e Tecnologia, \\
Universidade NOVA de Lisboa}
\date{	}

\newacronym{wanet}{WANET}{wireless ad-hoc network}
\newacronym{manet}{MANET}{mobile ad-hoc network}
\newacronym{vanet}{VANET}{vehicular ad-hoc network}
\newacronym{wsn}{WSN}{wireless sensor network}
\newacronym{p2p}{P2P}{peer-to-peer}
\newacronym{dtn}{DTN}{disruption tolerant network}
\newacronym{olsr}{OLSR}{optimized link state routing}
\newacronym{dsdv}{DSDV}{destination-sequenced distance-vector routing}
\newacronym{aodv}{AODV}{ad-hoc on-demand distance vector routing}
\newacronym{dsr}{DSR}{dynamic source routing protocol}
\newacronym{zrp}{ZRP}{zone routing protocol}
\newacronym{tdls}{TDLS}{tunneled direct link setup}
\newacronym{gpsr}{GPSR}{greedy perimeter stateless routing}
\newacronym{ap}{AP}{access point}
\newacronym{dht}{DHT}{distributed hash table}
\newacronym{ght}{GHT}{geographic hash table}
\newacronym{chr}{CHR}{cell hash routing}
\newacronym{udg}{UDG}{unit disk graph}
\newacronym{bs}{BS}{base station}
\newacronym{ps}{P/S}{publish/subscribe}
\newacronym{dnf}{DNF}{disjunctive normal form}
\newacronym{cnf}{CNF}{conjunctive normal form}
\newacronym{tps}{T-P/S}{time-aware publish/subscribe}
\newacronym{smr}{SMR}{state machine replication}
\newacronym{dcs}{DCS}{data-centric storage}
\newacronym{hcrp}{HCRP}{home cell refresh protocol}
\newacronym{prp}{PRP}{perimiter refresh protocol}
\newacronym{c2da}{C$^2$DA}{cell-by-cell destination aggregation}
\newacronym{mda}{MDA}{message destination aggregation}
\newacronym{ntp}{NTP}{network time protocol}
\newacronym{mec}{MEC}{mobile edge computing}
\newacronym{d2d}{D2D}{device-to-device}
\newacronym{nack}{NACK}{negative acknowledgement}
\newacronym{plsg}{PL/SG}{publish locally/subscribe globally}
\newacronym{rwp}{RWP}{random waypoint}
\newacronym{pds}{PDS}{peer data sharing}
\newacronym{cep}{CEP}{complex event processing}

\begin{document}

\maketitle

\begin{abstract}
Smart mobile devices are increasingly ubiquitous and
are the primary source of user-generated content,
and current communication infrastructures are
failing in keeping up
with the rising demand for the avid
sharing of such content.
To alleviate this problem and fully harness the
amount of resources currently available at the network edge,
\emph{mobile edge} paradigms started to emerge.
Though, application developers still struggle to
tap that potential at the edge due to the lack of adequate
communication and interaction abstractions.
Thus, we propose a high-level abstraction that can
be easily exploited by developers to design \emph{mobile edge}
applications focused on data dissemination.
In this paper, we propose \sys, a novel extended
topic-based, time-aware publish/subscribe system for networks
of mobile devices.
In \sys, time is a first order dimension.
Each subscription has an associated time frame, starting and
ending either in the future, present, or past.
Making the past available requires both subscriptions and publications to be
persistently stored.
We present the design of \sys and evaluate it using
simulation, discussing and characterizing the scenarios
best suited for its use.

\end{abstract}

\glsresetall
\section{Introduction}
\label{sec:intro}
Smart mobile devices, like smartphones and tablets, are
increasingly ubiquitous, and their increasing capabilities are turning
them into pocket-size personal computers.
%
%
The aggregate capacity of these mobile devices presents itself as 
massive computing and storage resource pools
that are still highly under-exploited~\cite{cisco:16}.
%
%

The pervasiveness of mobile devices makes them the primary
tool for generating and sharing all sorts of
content~\cite{cisco:16}, and
%
users expect to use such devices \emph{continuously} to both
access and share content like video or photos~\cite{pew}.
%
%
%
This usage pattern places a huge burden on network
infrastructures and cloud-based services alike, because
they have to accommodate high loads to support
continuous user activity, namely in sharing user-generated
content, \eg, in social gatherings such as sports events or
music concerts~\cite{cisco:16, Erman:2013}.
%
%
%
%

The typical alternative to sustain such high demand
is to set up special communication infrastructures just
for those events~\cite{verizon:16}.
Unfortunately, in some scenarios~(\eg, one-time events) it
might be logistically or financially nonviable to deploy
such infrastructures~\cite{baburajan:16}.
In others~(\eg, natural disasters, inhospitable locations),
infrastructures might not even exist or be impossible to
set up~\cite{Manoj:2007}.
%
%

To tackle such daunting scenarios and to fully exploit the amount
of resources that are now available at the network edge, novel
computational paradigms, such as mobile edge computing~\cite{7879258}
and mobile edge cloud~\cite{drolia:13}, have emerged.
%
%
%
Though, it is still challenging for application developers
to easily tap this potential, namely due to the lack of adequate
communication and interaction abstractions.
%
%

Recent research~\cite{serval,meshkit,teofilo:mobiquitous:2017,rodrigues:2016}
is leveraging on \gls*{d2d} communication to provide peer-to-peer ad-hoc
communication among mobile devices at the network edge.
These \gls*{d2d} mechanisms can complement and work side-by-side with
typical communication infrastructures~\cite{silva:mecc:2016}.
%
However, they still lack
adequate abstractions for developers, forcing them to
reason and address many issues that are related with the
mobile and wireless nature of the execution environment, which
is not only hard and time-consuming, but also error-prone.
%
%

In this work, we aim at providing a high-level abstraction
that can be readily exploited by developers to design ``mobile edge''
applications focused on the dissemination of data among
nearby mobile devices.
For that purpose, we focus on the \gls*{ps} paradigm.
%
%
This simple communication abstraction provides full decoupling in time,
space, and synchronization between publishers and
subscribers~\cite{Eugster:2003},
which facilitates loosely coupled, spontaneous interactions~(required
for this kind of dynamic and pervasive edge environments).
%
%

In this paper, we propose \sys, a novel extended
topic-based, \gls*{tps} system for wireless
networks of mobile devices, working as a data dissemination/storage
service at the network edge.
%
The extended topic-based feature stems from the fact that
\sys's subscriptions support \emph{arbitrary} propositional
logic formulas~(using topics as literals), whereas typical
topic-based systems only allow one topic per subscription.
%
%

In the scenarios we are addressing~(\eg, sports games), 
individual moments are
intrinsically tied by time relations~(\eg, the
goal in the first half of the football game).
Also, when consuming contents, users are often interested
in events that have these associated time
references~(\eg, find pictures of the
goalkeeper's save in the second half of the game).
Accordingly, \sys considers \emph{time} to be a first
order dimension and we define it as a \gls*{tps} system,
whose data is persistently stored, and 
where subscriptions include a time frame that define
its active time-span, either in the future, in the
present, or in the \emph{past}.
%
%
As a result, the target  publication space  is 
confined  to publications that
happen(ed) within the specified time frame, which 
effectively provides the full time decoupling of
the \gls*{ps} paradigm.
To the best of our knowledge, \sys is the first system
to expose, in this context, a \gls*{ps} interface with
support for subscriptions within a time scope that can
reside in the past.
%
%

We present two different materializations of \sys.
The first one, \sys{}-PL/SG, is a simplistic approach using flooding.
The second more intricate one, \sys{}-DCS, is inspired by the fact that
geographical positions have a close relation to topology in
wireless networks, and follows a \gls*{dcs} approach using a
\gls*{ght} as a storage substrate.
%
%

In summary, the contributions of this paper are the following:
\begin{inparaenum}[i)]
	\item the concept of a \gls*{ps} system with intrinsic
	time-awareness, requiring persistent subscriptions and
	publications~(\S\ref{sec:tps});
	\item the design of \sys, a novel extended topic-based
	\gls*{tps} system~(\S\ref{sec:overview}) and our two
	materializations of this proposal~(\S\ref{sec:plsg}
	and~\S\ref{sec:dcs}); and
	\item the characterization of the scenarios best
	suited for the use of the proposed solutions, using
	simulation~(\S\ref{sec:eval}).
\end{inparaenum}
%
%

\section{Time-Aware Publish/Subscribe}
\label{sec:tps}
This time-awareness concept stems from the facts that each
publication is timestamped, and each subscription is active
within a specific time frame.
%
Thus, even if a subscription's time frame lays partially or
entirely in the past, the subscriber will still be notified about~(past)
publications contained within that time frame.

A \gls*{tps} system offers the usual operations of a regular
\gls*{ps} system, namely publish data, subscribe to topics,
cancel a subscription, and retrieve previously published data.
To support subscriptions with a time span
in the past, published data must be persistently
stored within the system.
Accordingly, the \gls*{tps} interface also supplies an
operation to \emph{unpublish} data, deleting it from storage.

\subsection{Publishing Data}
\label{ssec:pub}
A data object is the basic unit of work and is
seen as an opaque set of bytes.
Every object has some associated metadata that
consists in a tuple $$\langle id_{\mathrm{obj}}, T, s,
ts^{\mathrm{pub}}, id_{\mathrm{owner}} \rangle$$, where:
\begin{inparadesc}
	\item $id_{\mathrm{obj}}$ is the object's identifier;
	\item $T$ is a set of tags or keywords related to
	the object, \eg, hashtags used in social networks;
	\item $s$ is a summary of the object,
	\eg, a thumbnail of an image;
	\item $ts^{\mathrm{pub}}$ is the object's publication
	timestamp; and
	\item $id_{\mathrm{owner}}$ is the publisher's
	node identifier.
\end{inparadesc}

The system-wide unique \emph{object key} is composed of
the object's identifier and the publisher's node
identifier, \ie,  it is in fact a pair
$\langle id_{\mathrm{obj}}, id_{\mathrm{owner}} \rangle$.
This enables different nodes to publish objects under
the same identifier, turning the object's identifier into a
sort of a domain key, hence avoiding
name collisions among nodes.

Tags are used as topics for subscriptions.
Tagging is a flexible annotation scheme,
\eg, by adding the publisher's node
identifier to the tags of its published objects, an application
can easily enable the retrieval of all the
objects published by a certain node/user.

\subsection{Removing Data}
\label{ssec:unpub}
The unpublish operation revokes a published object, making it inaccessible
to future subscriptions.
%
%
Note that a subscription targeting the past
will not see unpublished objects, even if these were 
initially available in the subscription's time frame.

\subsection{Subscribing}
\label{ssec:sub}
With time as a first order dimension, a subscription
consists in a tuple $$\langle id_{\mathrm{sub}}, q,
ts^{\mathrm{s}}, ts^{\mathrm{e}}, id_{\mathrm{owner}} \rangle$$, where:
\begin{inparadesc}
	\item $id_{\mathrm{sub}}$ is the subscription's identifier;
	\item $q$ denotes the query that defines which tags are
	relevant to the subscription;
	\item $ts^{\mathrm{s}}$ is the timestamp defining when the
	subscription's time frame starts;
	\item $ts^{\mathrm{e}}$ is the timestamp defining when the
	subscription expires; and
	\item $id_{\mathrm{owner}}$ is the subscriber's node identifier.
\end{inparadesc}

The query is a formula in propositional logic where
literals are tags associated with published
objects~(\eg, `$A \, \& \, (B \, | \, C)$' captures objects tagged
with $A$ and at least one of $B$ or $C$).

The $ts^{\mathrm{s}}$ and $ts^{\mathrm{e}}$ timestamps
specify the time frame in which the
subscription is valid, where the special value $\bot$
represents, respectively, the times at which the system
started and stopped to exist.
%
This allows the subscriber to specify any time frame that
might be relevant to the subscription.
For instance, assuming a subscription is submitted at
time $t$: $ts^{\mathrm{s}} = \bot$ and
$ts^{\mathrm{e}} = t$ refers to events that happened
before the subscription; $ts^{\mathrm{s}} = t$ and
$ts^{\mathrm{e}} = \bot$ refers to events 
after or concurrent with the subscription;
and $ts^{\mathrm{s}} = ts^{\mathrm{e}} = \bot$ refers to
events that occurred at any time.
Notice that these parameters can also take concrete
timestamp values.

Regarding notifications, there are two situations that can
trigger the notification of subscribers:
upon a publication, the detection that the object being
published matches existent subscriptions;
and upon receiving a subscription that spans into the past, the
detection that previously published objects match
this new subscription.
%
In both cases notifications are sent to the respective
nodes carrying the metadata of the matching objects.
The objects' metadata is the only information
given to the subscribers for them to decide if objects
are relevant enough for retrieval.

The unsubscribe operation revokes
a subscription before it naturally expires after its
end timestamp, $ts^{\mathrm{e}}$.

\subsection{Retrieving Data}
\label{ssec:down}
Received notifications must be acted upon, and may either be discarded, trigger an immediate download, or be stored by the application and acted upon later.

In order to retrieve some published object, the retriever
needs to know where to get that specific object from.
This is a classical resource discovery problem.
In \sys, the minimum amount of information needed to get an
object is its key~(the pair $\langle id_{\mathrm{obj}},
id_{\mathrm{owner}} \rangle$).
%
%
%
%
%

\glsresetall
\section{The Many Leaves of \sys}
\label{sec:overview}
Now, we present a general overview of \sys's design, providing
a \gls*{tps} interface.

\subsection{Use Cases}
As a \gls*{ps} system, \sys provides a generic data dissemination service.
We argue that \sys fits perfectly in scenarios where
crowds are gathered, using their mobile devices to collect
data~(\eg, photos, video, text) and share it with
people in their vicinity, akin to social networks~\cite{seada2006social}.

Consider, for instance, a scenario where spectators in
different parts of a football stadium may share their
views of the game through self-generated~(multimedia) content.
In this case, spectators would be able to see key moments
of the game from multiple viewpoints, including those of the
spectators in key locations or closer to the field.

This kind of augmented user experience is already being
explored using venues' fixed communication
infrastructures~\cite{yinzcam}, which may be subjected to
overload conditions and failures~(\eg, power outages~\cite{Erman:2013}).
In turn, the \textit{mobile edge} paradigms and the advances in
\gls*{d2d} communication offer the possibility
to provide such enriched user experience with a negligible cost
for infrastructure managers, while at the
same time working to alleviate the load on those infrastructures.

\subsection{System Model}
We consider a classical asynchronous model comprised
of~$\mathrm{\Pi} = \{n_1, \dots, n_k\}$ nodes, with
no mobility restrictions, other than those imposed by the venue
they are in, and the natural speed limits of humans. 
We do not assume any specific radio technology.
Nodes communicate by exchanging messages
through a wireless medium~(\eg, Bluetooth, Wi-Fi ad-hoc,
Wi-Fi Direct), and have no access to any
form of shared memory.
Nonetheless, nodes should be able to establish
communication channels with~(all) their one-hop neighbors.
We also consider the classical crash-stop failure model,
whereby nodes can fail by crashing but do not behave maliciously.

Published object data is considered immutable.
We do not consider security or access control concerns thus, only
publicly sharable data is published~(\eg, as in social networks).
Due to the unreliable nature of wireless communication
mediums, \sys notifies subscribers of all relevant published
data as completely and faithfully as possible, \ie, missing
some notifications is permitted because applications are not
expected to be mission-critical.

Each node has a globally unique identifier and can
determine its geographical position, either
through GPS or other means~\cite{Rai:2012}.
Thus, nodes can be aware if they are moving or stationary.
We also assume nodes' clocks to be
synchronized~(with a negligible skew).
Both these assumptions are reasonable since we target
mobile devices~(\eg, smartphones) and nowadays, even
low-end devices come equipped with GPS and synchronize
their clocks with the network providers, while other solutions
allows to effectively locate a user device even indoor 
(for instance through the monitoring of visible access points).

\subsection{Architecture}
In \sys, akin to~(flat) \gls*{p2p} systems, nodes are
functionally symmetric, sharing the same responsibilities
and having no particular roles.
There are no centralized or specialized components~(like
\gls*{p2p} super-peers or \gls*{ps} brokers), and each node can
be a publisher, a subscriber, or both.

\begin{figure}[tb]
	\centering
	\includegraphics[scale=.65]{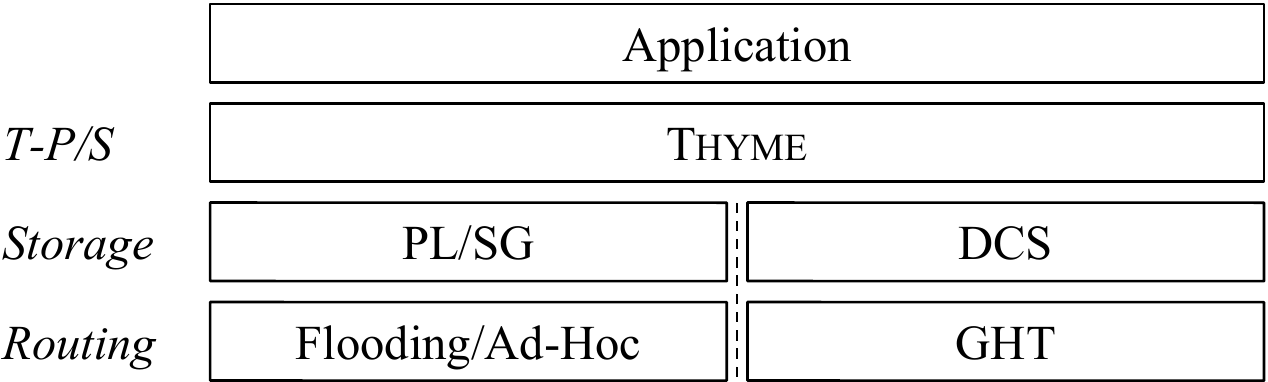}
	\caption{System overview.}
	\label{fig:sys-overview}
\end{figure}

\sys's design comprises three main layers, depicted in
Figure~\ref{fig:sys-overview}.
The bottom layer handles message routing.
The middle layer addresses \sys's time-awareness, which 
%
requires both subscriptions and
publications to be persistent.
Such requirement is handled by the middle, storage layer.
The top layer is \sys itself, providing the \gls*{tps} interface
for applications.

As illustrated in Figure~\ref{fig:sys-overview}, we
present two materializations for the two bottom
layers~(routing and storage):
\begin{inparaenum}[1)]
	\item one version uses a \gls*{plsg}
	approach, where~(un)publish operations are executed
	locally and~(un)subscribe operations are flooded to every
	node in the system~(see~\S\ref{sec:plsg});\label{ver:plsg}
	\item the other version follows a more intricate
	\gls*{dcs} approach using a \gls*{ght} for
	storage~(see~\S\ref{sec:dcs}).\label{ver:dcs}
\end{inparaenum}

\subsubsection{Routing}
In version~\ref{ver:plsg}~(\gls*{plsg}), the routing layer
provides flooding to the entire network~(using unreliable UDP broadcast),
%
%
and~(multi-hop) unicast using a typical ad-hoc routing
protocol~(\eg, DSDV, OLSR).
In version~\ref{ver:dcs}~(\gls*{dcs}), the routing layer
is materialized as a cluster-based \gls*{ght} for
wireless networks.
Thus, the routing infrastructure is based on the notion
of clusters or virtual nodes.
Physical space is divided into a grid, \ie, equally-sized
square-shaped cells, and all physical nodes within the
geographic boundaries of a cell collaboratively act
as a virtual node.
Messages are addressed to geographic locations, thus routed 
to the cell that contains the location of the message destination.
Messages addressed to a virtual node, \ie, a cell, are
delivered to all physical nodes that materialize
the virtual node managing the cell~(in a similar way
to~\cite{Leitao:2014}).

Wireless communication mediums are known to be subject
to many forms of interference; hence some messages may be
lost and not reach their final destinations.
However, as a design principle in both versions, we do not
provide any mechanisms to recover from lost messages on
the wireless medium, delegating this responsibility
to the upper layers~(abiding by the
end-to-end argument~\cite{endtoend}).

\subsubsection{Storage}
Since subscriptions can target the past,
publications have to be kept persistent in the system.
So, \sys's architecture conceptually fuses a
\gls*{ps} system with a storage system.

In the \gls*{plsg} approach, each node stores its
publications locally, while subscriptions are fully
replicated in every node of the system.
In the \gls*{dcs} approach, the use of
the underlying \gls*{ght} is two-fold:
\begin{inparaenum}[1)]
	\item virtual nodes~(\ie, cells) are
	used to store published content and subscriptions; and
	\item routing is exploited to match
	subscriptions and publications, \ie, cells act as
	virtual \gls*{ps} brokers.
\end{inparaenum}
%
%

\section{\sys-PL/SG}
\label{sec:plsg}
This materialization of \sys employs a simple
\gls*{plsg} approach~(version~\ref{ver:plsg}).
%
%
Publish and unpublish operations are entirely executed locally.
Thus, publications are only stored~(locally) by their owners.
%
%

On the other hand, subscribe and unsubscribe operations
are flooded and executed in every node of the system.
Hence, subscriptions are stored by every node, \ie, fully replicated.
%
Notifications are triggered in two situations:
upon a publication, the publisher node checks if the object
being published matches any of the subscriptions it
has stored;
and upon a subscription~(when
flooding the respective message), each node
that receives the subscription message checks if that new
subscription matches any of the node's previously
published objects.
%
%

Download operations request the desired objects directly
from the object owners, using the publishers' node identifiers
contained in the objects' metadata received in the notifications.
Here, the multi-hop unicast provided by the routing
layer is used to contact the object owners directly.
%
To recover from lost messages on the wireless medium, this
operation uses a retransmission mechanism.
After a configured amount of time has passed without receiving
a reply, the operation is retried.
If the~(configured) maximum number of retries is reached, the
operations fails with a timeout error code.
%
%

In this version, node mobility is handled in a completely transparent
way by the protocol used in the routing layer.
%
%

Since publications are only stored locally by the
object owners, this materialization does not guarantee
the persistence of publications once the publisher node fails.
%
%

For a node to join the system, it first broadcasts a join request.
To avoid replies from all the nodes that received the
join request, only some~(randomly selected) nodes will respond
with all the subscriptions they have locally stored.
The replies are also delayed a random amount of time~(in a
configured interval) to avoid message collisions in the wireless medium.
This join procedure also uses a retransmission mechanism,
similar to the download operation.
If the maximum number of retries is reached, the joining node
assumes it is alone in the network, and starts working as normal.
%
%

\section{\sys-DCS}
\label{sec:dcs}
In this version, \sys is materialized on top of a
cluster-based geographic routing layer~(version~\ref{ver:dcs}).
It follows a \gls*{dcs} approach making
use of the underlying \gls*{ght} to provide a
storage substrate.
This approach has two complementary aspects:
\begin{inparaenum}[1)]
	\item it provides topology-awareness by
	design; and
	\item it allows the inference of the location of
	relevant data to subscriptions, enabling access to such
	data using a location-aware strategy.
\end{inparaenum}
This materialization leverages heavily on the notion of
cell~(or virtual node) conveyed by its routing layer.

\subsection{Publishing Data}
\label{ssec:pub2}
When executing a publish operation, this approach
leverages on the cells conveyed by the underlying \gls*{ght}
and it indexes the object metadata by its tags $T$, \ie, the
cells resultant from hashing each tag in $T$ store the metadata.
The actual object is only stored in the publisher node and its
current cell~(see~\S\ref{ssec:rep}).
This ensures only the object's metadata is sent through the network.

\begin{figure*}[tb]
	\centering
	\includegraphics[scale=.54]{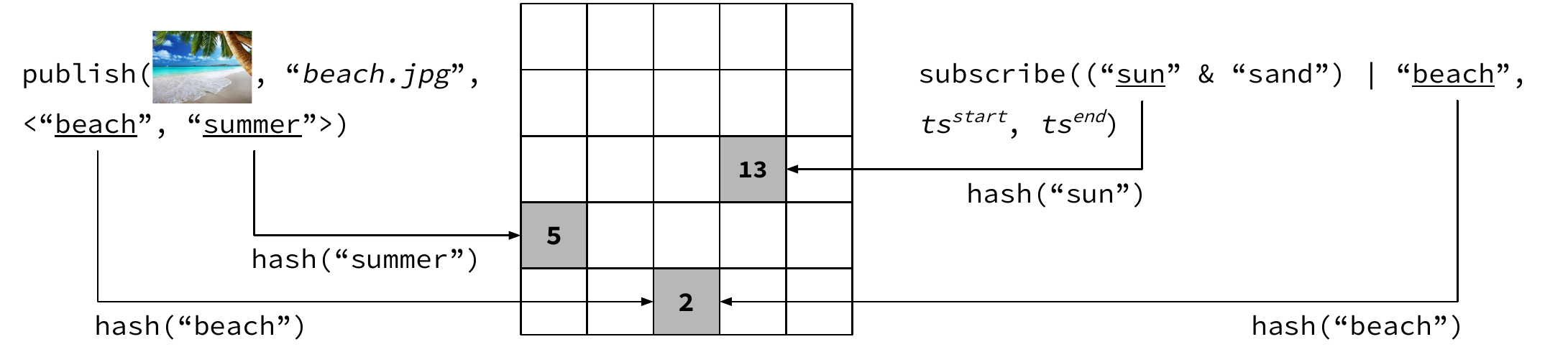}
	\caption{Publish and subscribe operations.
	The tags' hashing determines the cells responsible
	for managing the object metadata~(cells~2 and~5) and
	the subscription~(cells~2 and~13).
	If a subscription has overlapping tags with a
	publication~(and vice versa) it will also have
	overlapping~(responsible) cells, guaranteeing the
	matching and sending of notifications to the subscriber.}
	\label{fig:pubsub}
\end{figure*}

Figure~\ref{fig:pubsub} illustrates the publication of
a photo with identifier ``beach.jpg'', and tags ``beach''
and ``summer''.
The cells resultant from hashing each tag are responsible for
managing the object's metadata and checking
if subscriptions match this publication.

\subsection{Replication}
\label{ssec:rep}
Since we target dynamic and pervasive edge environments,
in order to provide data availability and tolerance to
churn, this materialization provides two replication mechanisms.

\subsubsection{Active Replication}
\label{sssec:act_rep}
It takes advantage of the virtual nodes
provided by the cluster-based \gls*{ght}.
Upon publishing, an object is disseminated inside the
publisher's cell.
Onwards, every node inside the cell should
be able to reply to download requests for that object.
This ensures tolerance to churn and guarantees
that published content will remain in the system
even if publishers leave.

\subsubsection{Passive Replication}
\label{sssec:pass_rep}
It leverages on the nodes that
already downloaded an object to provide more replicas of
that same object scattered in the network, increasing
data availability.
At the same time, it offers a list of multiple locations
from where the object may be downloaded.

\subsubsection{Replication List}
To take advantage of both these replication
mechanisms, the system needs to keep track of the
whereabouts of each object replica.
This bookkeeping is done by listing an object's replicas
location in its metadata, in what we call replication lists,
$L_{\mathrm{rep}}$~(a list of pairs $\langle id_{node}, cell_{node} \rangle$).
Thus, in this case, the object metadata consists of a
tuple $\langle id_{\mathrm{obj}}, T, s, ts^{\mathrm{pub}},
id_{\mathrm{owner}}, L_{\mathrm{rep}} \rangle$.

Since nodes can move, their location may change over time.
Hence, after a node stabilizes in a~(new) cell, it must
update its location for the
passive replicas of the objects it holds.

\subsection{Removing Data}
\label{ssec:unpub2}
In the unpublish operation, the object metadata indexed
by the object's tags is removed from the responsible cells.
However, while the active replicas are also explicitly
removed, the same does not happen to the passive ones.

\subsection{Subscribing}
\label{ssec:sub2}
Subscriptions are extended with the location of the
subscriber node, $cell_{\mathrm{owner}}$.
Thus, a subscription consists of a tuple
$$\langle id_{\mathrm{sub}}, q, ts^{\mathrm{s}}, ts^{\mathrm{e}},
id_{\mathrm{owner}}, cell_{\mathrm{owner}} \rangle$$.
The subscriber's cell address is required because
the geographic routing layer only routes messages to
geographical positions.
Hence, there is the need to know the cell where to send
notifications to.
This information needs to be updated every time the
subscriber node changes its cell.
%
%

\subsubsection{Divide and Conquer}
\label{sssec:div_conquer}
Here, we leverage on the fact that every
propositional logic formula has an equivalent in
\gls*{dnf}.
Thus, we employ a divide and conquer strategy of breaking the
disjunction into its individual conjunctive clauses, and
evaluate each one separately.
For a match to occur, it suffices that one of the conjunctions
evaluates to true.
The use of \gls*{dnf} enables load balancing when checking
for a match between a publication and a subscription,
since the work of verifying a match can be split among
different cells/nodes, each evaluating only one of the
query's conjunctions.
Additionally, it minimizes the amount of information
transmitted to the responsible cells, since each subscription
message only needs to carry the respective conjunction.

For each conjunction, we select a random \emph{non-negated}
literal as its key.
The result of hashing that literal determines the cell
where to send that part of the query.
That cell becomes a~(virtual) broker for the subscription, and the
nodes in the cell are responsible for checking for
publications matching the subscription, and notifying the
subscriber if need be.
Figure~\ref{fig:pubsub} depicts a subscription of a query with
two conjunctions.
For each, one of its~(non-negated) literals is chosen as
its key, and determine the cells that will become the virtual
brokers for the subscription.

\subsubsection{Notifications}
%
Upon a publication, the cells
indexing the object metadata by its tags are responsible for
checking if the publication being indexed matches any
existent subscriptions stored locally.
Upon a subscription,  the cells indexing the
subscription by its keys are responsible for checking
if the locally stored objects match the new subscription.

\subsubsection{Excess of Past Notifications}
When issuing a subscription for some time in the past,
and for a tag with many previously published objects,
the subscriber will be flooded by a large
amount of notifications.
Besides flooding the subscriber with lots of notifications,
this also implies lots of communication.

To attenuate this problem, when subscribing for some time
in the past, the subscriber is only notified about $n$
matching objects from a total of $x$ objects.
Then, if interested, the subscriber can request more
matching objects.
Receiving the notifications in expressly requested batches.
All the subsequent publications matching the subscription
will be notified as usual.

\subsubsection{Moving Subscribers}
\label{sssec:move-subs}
When a subscriber moves to a different cell, it must update its
location for every active subscription it owns.
Notifications sent to subscribers on the move may
never reach their destination.
In such cases, the underlying routing layer returns
\glspl*{nack}~(see~\S\ref{ssec:nacks}) for messages addressed
to individual nodes that could not be delivered.
\glspl*{nack} are used to convey that a node is no longer
in  its supposed cell, which may be caused  by movement
or node failure.
Node movement will be detected through the subscriber's
location update.
In such case, we can re-send the notifications that were
not previously delivered.
Otherwise, we can simply stop sending notifications.

\subsubsection{Unsubscribing}
For each subscription, the subscriber keeps the list of
subscription keys.
When executing an unsubscribe operation, unsubscribe messages
are sent to the cells determined by the hashing of each key.

\subsection{Retrieving Data}
\label{ssec:down2}
Download operations leverage the replication mechanisms in
order to have multiple locations from where to retrieve
an object.
From all the locations in the replication list, the node
chooses the closest one to itself, and sends a download
request for the desired object.
If a negative reply is received, the requester proceeds
and tries to download the object from the next location
in the replication list~(until no more options are
available, or a maximum number of retries is reached).
In case the last retry is reached, it will always try to
download the object from the cell actively replicating
it~(if it was not already tried), because it offers
higher chances of success comparing to every other replica.

The use of geographical routing makes it easier for nodes
to make hints on which ``download locations''
are better~(\ie, closer).
The geographical approach makes possible to use a more
concrete metric for distance in the network than the
number of hops.
This approach reduces the distance data has to
travel in the network, when retrieving an object.

\subsection{Per Operation Retransmission Mechanism}
\label{ssec:retrans}
Similarly to the first materialization of \sys~(\S\ref{sec:plsg}), to
address the problem of collisions and interferences on the
wireless medium, this materialization also employs a per
operation retransmission mechanism~(for all the five \sys operations).
In this materialization, one operation may result in the
sending of multiple messages to different cells~(\eg, publishing
an object with two tags, as in Figure~\ref{fig:pubsub},
results in sending publish messages to two different cells).
Thus, after a configured amount of time without receiving the
expected replies~(one from each cell), the message is
retransmitted~(until a configured maximum number of retries),
but only for the cells that did not send a reply back.
%
%
%


\subsection{Joining the System}
\label{ssec:join}
Instead of sending a broadcast, like the \gls*{plsg} version,
in this materialization, a node joining the system waits a
configurable amount of time, listening for the periodic
beacons sent by all nodes.
During that waiting time, if the node listens to some
beacon sent by another node in its cell, the sender of
that beacon is used as the entry point to the system, and
exchange a join request and respective
reply~(with all the cell state).

This join procedure also uses a retransmission mechanism.
After sending a join request, if a reply is not received
within a configurable amount of time, the joining node
starts another round of waiting for the periodic beacons.
If the maximum number of retries is reached, it
assumes it is alone in the cell, and starts working as normal.
%
%

%

\subsection{Cluster-Based Geographic Routing}
\label{ssec:ght}
In a geographical distributed hash table, nodes know their
positions and keys are hashed into that same
domain.
The node responsible for a key is the one closest to the key's
geographical position.

Inspired by works from both wired~\cite{Leitao:2014,6623636}
and wireless~\cite{1303252,Araujo:2005} settings, we adopt a
cluster-based approach.
This approach conveys the notion of virtual nodes or cells.
Space is divided into a grid~(like in Figure~\ref{fig:pubsub})
and all nodes inside each cell act collaboratively as one.
%
%


\subsubsection{Routing}
Our routing scheme is very similar to the ones
used in~\cite{Araujo:2005,Ratnasamy:2002}.
Routing is done on top of these virtual nodes, \ie, at cell-level, using a
geographic routing protocol---a variation of the \gls*{gpsr}
protocol~\cite{Karp:2000}.
\Gls*{gpsr} makes greedy routing decisions, forwarding messages to the
next neighbor geographically closer to the message destination.
When such strategy is not possible, the algorithm resorts to a recovery
mode that forwards messages around the voids in the network.
For forwarding messages from cell to cell, we use unicast in
order to take advantage
of the~(per hop) MAC-level retransmission mechanism.

The interface exposed by this routing layer provides operations to
route messages to an individual node~(within a specific cell) and to
broadcast messages within the context of a single cell, besides
providing a routing mechanism between cells.

The one-hop broadcast is used as a neighbor discovery
service~(transmitting periodic beacons with the node's
current cell), and
as the intra-cell communication primitive.
Since broadcasts are not acknowledged at the MAC-level, they
are more likely to fail, which makes it a best-effort
communication primitive.

\subsubsection{Dynamic Cell Population}
\label{sssec:dyn_cell_pop}
Since it is impossible to ensure every cell has at least one
node, some keys may be left without nodes to manage them,
\ie, cells may be empty.
We address this in the same way as~\cite{Araujo:2005,Ratnasamy:2002},
forcing keys to take an entire loop around the empty cells,
stopping in the cell closest to the supposed destination~(which
becomes a proxy of the key's destination cell).

This raises another problem when nodes populate previously empty cells,
or leave the system and make some cell empty.
A cell leaving the network delivers all its keys to its proxy
cell.
An entering cell needs to receive its keys from its
proxy cell, and also all the keys of empty cells for
which it act as a proxy.


\subsubsection{Mobility}
Concerning mobility, we argue that moving nodes render routing
information volatile.
Thus, only nodes that are stationary actively participate
in the routing of messages.
Since our target scenarios have mild mobility patterns~(\ie, nodes
do not move constantly, and some might not even move during the
entire event), only stationary nodes form the \gls*{ght}.
When a node starts to move, and leaves its current cell, it
stops participating in the routing
protocol~(\ie, it stops forwarding messages).
It resumes the protocol when it detects itself as being stationary,
by joining the local cell.
While moving, nodes still process received periodic beacons,
allowing them to keep communicating with the \gls*{ght}.

\subsubsection{Negative Acknowledgements}
\label{ssec:nacks}
Nodes are not individually addressable, but by knowing its
current cell, a message can be sent to a node in that specific cell.
To allow the upper layers to react to a node's failure or migration from one cell to another, the routing layer replies with a \gls*{nack} to a
message source node, when a message addressed to an individual node
could not be delivered.
%
%
%
%

\subsubsection{Cell-by-Cell Destination Aggregation}
For messages that are to be delivered to multiple
destinations~(\eg, notifications), we optimized
our routing scheme by only propagating a single 
message to those destination. This message is only duplicated when
strictly required, which happens when the message's next hop for
different destinations is not the same.
This contributes to reduce energy consumption and the
occupancy of the wireless medium.
%
%

%
%


\section{Evaluation}
\label{sec:eval}
Our evaluation seeks to answer the following questions:
\begin{itemize}
	\item Which are the trade-offs provided by each version of \sys?
	\item How does each materialization deals with churn?
	\item How do they react to node mobility?
\end{itemize}
Each data point reports the average of~5 randomly
generated network topologies, each independently
run~3 times, making a total of~15 runs per data point.

\subsection{Implementation}
A meaningful real world evaluation would require a significant
number of mobile devices.
Not having access to such infrastructure, we resort to
simulation and implement both materializations of \sys in
the ns-3 network simulator~\cite{Riley2010}.
We use ns-3.27 and nodes communicate through WiFi ad-hoc~(using UDP).
The \gls*{plsg} version uses DSDV as the routing protocol.
%
%
%

In the \gls*{dcs} version of \sys, when a cell becomes
empty~(\ie, when all nodes leave a cell), all the data
stored in it needs to be transfered to another cell.
The reverse happens when an empty cell becomes
populated~(\S\ref{sssec:dyn_cell_pop}).
Currently, we do not implement such mechanism thus, in our
experiments, cell population is static (\ie, populated and
empty cells will remain as such throughout the system's lifetime).
This poses some limitations regarding node mobility and churn
in the \gls*{dcs} version.
A node cannot leave its cell if it is the only one in
it~(but it can still move inside its cell).
Also, nodes can only move among populated cells.
%
%

%
%
%
%
%
%

%

%

\subsection{Simulator Setup and Traces}
Unless stated otherwise, all non-mentioned parameters were left
with the simulator's default values.
We used Wi-Fi 802.11g configured with a constant rate
manager and a data rate of~6 Mbps.
The RTS/CTS threshold was~1500 bytes.

In order to mimic a realistic scenario, we emulate an application
similar to a online social network on top of \sys, and we use
tweets as user publications.
We generate operation trace files with the scenarios
behaviors, \ie, all the operations to be issued.

We crawled the tweets issued during the~2016 UEFA European
Championship final between Portugal and France.
The tweets were used as user publications, where:
the tweet id was used as the object identifier;
the text was used as the object's data;
the tweet timestamp was used as the object publication time; and
the hashtags were used as the object's tags.
For the traces' generation, the top-k most active users
were chosen, and every other operations were generated
from that, using exponential distributions configured with
different~$\lambda$ values~(\ie, rates).
Subscriptions were generated taking into account the
tags of the published objects, and
%
the top~$60\;\%$ of the most popular
tags were for the subscriptions' queries~(for
simplicity, each subscription subscribed to only one tag chosen at random).
Subscriptions were generated in two forms: in the
past~(with $ts^{\mathrm{s}} = ts^{\mathrm{e}} = \bot$); and in the
future~(with $ts^{\mathrm{s}} = now$ and $ts^{\mathrm{e}} = \bot$).
Subscriptions in the past where generated with a probability of~60\%.
During the first half of the game, subscriptions for each user were
generated with a rate of~3 subscriptions per hour.
During the rest of the event, subscriptions were generated with
a rate of~1 per hour.
Unpublications and unsubscriptions, which are expected to be rare
operations, were generated with a rate of~0.5 and~0.2 per node per hour,
respectively, both only during the second half of the game.
We crawled a total of~3 hours, starting at~20:00 2016-07-10.
To make simulation execution more timely, we compressed the~3 hours of the game
into~10 minutes of simulated time.

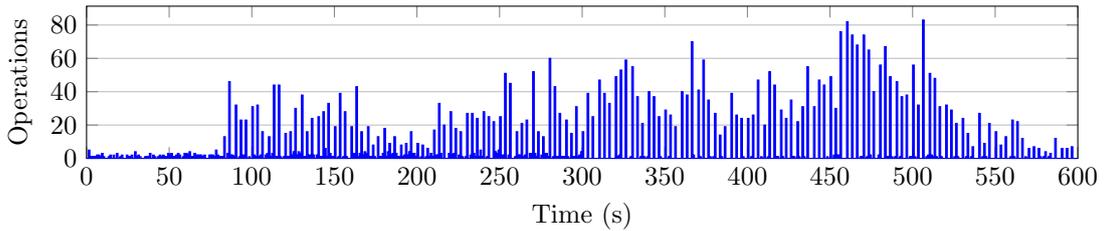
\begin{figure}[tb]
\centering
\begin{tikzpicture}
\begin{axis}[
width=0.98\textwidth,
height=3.6cm,
xmin=0, xmax=600,
ymin=0,
xlabel=Time (s),
ylabel=Operations,
ymajorgrids,
]
\addplot[ybar interval, fill, color=blue] coordinates {
	(0,0)
	(1,5)
	(2,1)
	(3,1)
	(4,1)
	(5,1)
	(6,2)
	(7,2)
	(8,1)
	(9,3)
	(10,1)
	(11,0)
	(12,0)
	(13,1)
	(14,2)
	(15,0)
	(16,2)
	(17,0)
	(18,3)
	(19,0)
	(20,1)
	(21,2)
	(22,0)
	(23,0)
	(24,2)
	(25,1)
	(26,1)
	(27,2)
	(28,0)
	(29,4)
	(30,1)
	(31,2)
	(32,1)
	(33,0)
	(34,0)
	(35,1)
	(36,1)
	(37,1)
	(38,3)
	(39,0)
	(40,2)
	(41,1)
	(42,0)
	(43,1)
	(44,2)
	(45,1)
	(46,2)
	(47,1)
	(48,1)
	(49,3)
	(50,2)
	(51,3)
	(52,1)
	(53,1)
	(54,2)
	(55,3)
	(56,2)
	(57,0)
	(58,1)
	(59,3)
	(60,3)
	(61,2)
	(62,4)
	(63,1)
	(64,1)
	(65,3)
	(66,2)
	(67,2)
	(68,2)
	(69,2)
	(70,1)
	(71,2)
	(72,0)
	(73,0)
	(74,2)
	(75,2)
	(76,2)
	(77,1)
	(78,5)
	(79,2)
	(80,1)
	(81,1)
	(82,1)
	(83,13)
	(84,0)
	(85,3)
	(86,46)
	(87,2)
	(88,2)
	(89,1)
	(90,32)
	(91,0)
	(92,1)
	(93,23)
	(94,2)
	(95,2)
	(96,23)
	(97,0)
	(98,1)
	(99,0)
	(100,31)
	(101,2)
	(102,1)
	(103,32)
	(104,1)
	(105,2)
	(106,16)
	(107,0)
	(108,2)
	(109,3)
	(110,13)
	(111,1)
	(112,3)
	(113,44)
	(114,1)
	(115,1)
	(116,44)
	(117,1)
	(118,0)
	(119,1)
	(120,15)
	(121,1)
	(122,1)
	(123,16)
	(124,2)
	(125,4)
	(126,30)
	(127,2)
	(128,4)
	(129,3)
	(130,38)
	(131,1)
	(132,1)
	(133,16)
	(134,2)
	(135,2)
	(136,24)
	(137,2)
	(138,0)
	(139,2)
	(140,25)
	(141,1)
	(142,1)
	(143,28)
	(144,6)
	(145,1)
	(146,33)
	(147,3)
	(148,1)
	(149,1)
	(150,19)
	(151,0)
	(152,1)
	(153,39)
	(154,0)
	(155,1)
	(156,28)
	(157,1)
	(158,0)
	(159,0)
	(160,19)
	(161,3)
	(162,3)
	(163,43)
	(164,1)
	(165,3)
	(166,16)
	(167,2)
	(168,4)
	(169,3)
	(170,19)
	(171,1)
	(172,1)
	(173,8)
	(174,1)
	(175,0)
	(176,13)
	(177,1)
	(178,1)
	(179,3)
	(180,18)
	(181,1)
	(182,4)
	(183,9)
	(184,2)
	(185,3)
	(186,13)
	(187,3)
	(188,1)
	(189,1)
	(190,8)
	(191,1)
	(192,2)
	(193,9)
	(194,2)
	(195,4)
	(196,16)
	(197,3)
	(198,3)
	(199,1)
	(200,9)
	(201,0)
	(202,1)
	(203,8)
	(204,0)
	(205,1)
	(206,6)
	(207,1)
	(208,3)
	(209,1)
	(210,17)
	(211,2)
	(212,2)
	(213,33)
	(214,3)
	(215,0)
	(216,20)
	(217,0)
	(218,0)
	(219,3)
	(220,28)
	(221,2)
	(222,3)
	(223,18)
	(224,0)
	(225,2)
	(226,16)
	(227,1)
	(228,0)
	(229,1)
	(230,27)
	(231,0)
	(232,2)
	(233,27)
	(234,3)
	(235,1)
	(236,24)
	(237,1)
	(238,5)
	(239,1)
	(240,28)
	(241,0)
	(242,1)
	(243,25)
	(244,0)
	(245,0)
	(246,22)
	(247,4)
	(248,3)
	(249,5)
	(250,25)
	(251,2)
	(252,2)
	(253,51)
	(254,0)
	(255,2)
	(256,45)
	(257,0)
	(258,0)
	(259,1)
	(260,16)
	(261,2)
	(262,2)
	(263,21)
	(264,1)
	(265,2)
	(266,23)
	(267,0)
	(268,3)
	(269,3)
	(270,52)
	(271,3)
	(272,1)
	(273,16)
	(274,0)
	(275,2)
	(276,13)
	(277,3)
	(278,3)
	(279,1)
	(280,60)
	(281,0)
	(282,1)
	(283,43)
	(284,1)
	(285,1)
	(286,27)
	(287,1)
	(288,1)
	(289,1)
	(290,23)
	(291,1)
	(292,0)
	(293,15)
	(294,1)
	(295,1)
	(296,31)
	(297,1)
	(298,2)
	(299,4)
	(300,16)
	(301,0)
	(302,0)
	(303,39)
	(304,0)
	(305,0)
	(306,25)
	(307,0)
	(308,0)
	(309,0)
	(310,47)
	(311,0)
	(312,0)
	(313,39)
	(314,0)
	(315,0)
	(316,33)
	(317,0)
	(318,0)
	(319,0)
	(320,49)
	(321,1)
	(322,2)
	(323,53)
	(324,0)
	(325,0)
	(326,59)
	(327,0)
	(328,0)
	(329,1)
	(330,55)
	(331,0)
	(332,0)
	(333,37)
	(334,0)
	(335,1)
	(336,21)
	(337,1)
	(338,0)
	(339,1)
	(340,40)
	(341,0)
	(342,0)
	(343,37)
	(344,0)
	(345,0)
	(346,25)
	(347,1)
	(348,0)
	(349,0)
	(350,29)
	(351,0)
	(352,1)
	(353,26)
	(354,0)
	(355,0)
	(356,19)
	(357,0)
	(358,0)
	(359,0)
	(360,40)
	(361,0)
	(362,0)
	(363,38)
	(364,0)
	(365,0)
	(366,70)
	(367,2)
	(368,0)
	(369,0)
	(370,41)
	(371,1)
	(372,0)
	(373,59)
	(374,0)
	(375,0)
	(376,35)
	(377,1)
	(378,0)
	(379,1)
	(380,27)
	(381,0)
	(382,0)
	(383,14)
	(384,1)
	(385,0)
	(386,19)
	(387,0)
	(388,0)
	(389,0)
	(390,39)
	(391,0)
	(392,0)
	(393,26)
	(394,0)
	(395,0)
	(396,24)
	(397,1)
	(398,0)
	(399,0)
	(400,24)
	(401,0)
	(402,1)
	(403,26)
	(404,0)
	(405,0)
	(406,47)
	(407,0)
	(408,1)
	(409,0)
	(410,20)
	(411,0)
	(412,1)
	(413,52)
	(414,0)
	(415,0)
	(416,44)
	(417,1)
	(418,1)
	(419,0)
	(420,29)
	(421,1)
	(422,0)
	(423,24)
	(424,0)
	(425,0)
	(426,35)
	(427,1)
	(428,1)
	(429,0)
	(430,22)
	(431,1)
	(432,0)
	(433,31)
	(434,1)
	(435,0)
	(436,55)
	(437,0)
	(438,0)
	(439,0)
	(440,31)
	(441,0)
	(442,0)
	(443,47)
	(444,0)
	(445,0)
	(446,44)
	(447,0)
	(448,1)
	(449,0)
	(450,49)
	(451,0)
	(452,2)
	(453,30)
	(454,2)
	(455,2)
	(456,76)
	(457,0)
	(458,0)
	(459,0)
	(460,82)
	(461,0)
	(462,0)
	(463,74)
	(464,0)
	(465,0)
	(466,68)
	(467,0)
	(468,1)
	(469,1)
	(470,74)
	(471,0)
	(472,1)
	(473,65)
	(474,2)
	(475,0)
	(476,40)
	(477,0)
	(478,0)
	(479,2)
	(480,56)
	(481,0)
	(482,0)
	(483,67)
	(484,1)
	(485,1)
	(486,49)
	(487,1)
	(488,0)
	(489,1)
	(490,46)
	(491,0)
	(492,1)
	(493,37)
	(494,0)
	(495,0)
	(496,38)
	(497,0)
	(498,1)
	(499,1)
	(500,56)
	(501,1)
	(502,0)
	(503,32)
	(504,0)
	(505,0)
	(506,83)
	(507,0)
	(508,1)
	(509,2)
	(510,51)
	(511,2)
	(512,1)
	(513,48)
	(514,1)
	(515,0)
	(516,31)
	(517,1)
	(518,0)
	(519,0)
	(520,32)
	(521,0)
	(522,0)
	(523,29)
	(524,0)
	(525,1)
	(526,21)
	(527,1)
	(528,0)
	(529,0)
	(530,24)
	(531,0)
	(532,1)
	(533,15)
	(534,0)
	(535,1)
	(536,7)
	(537,0)
	(538,0)
	(539,0)
	(540,27)
	(541,0)
	(542,1)
	(543,9)
	(544,1)
	(545,0)
	(546,21)
	(547,0)
	(548,0)
	(549,0)
	(550,16)
	(551,0)
	(552,0)
	(553,8)
	(554,0)
	(555,0)
	(556,13)
	(557,0)
	(558,0)
	(559,1)
	(560,23)
	(561,1)
	(562,0)
	(563,22)
	(564,0)
	(565,0)
	(566,12)
	(567,0)
	(568,1)
	(569,0)
	(570,6)
	(571,0)
	(572,0)
	(573,7)
	(574,0)
	(575,0)
	(576,6)
	(577,1)
	(578,0)
	(579,1)
	(580,4)
	(581,0)
	(582,1)
	(583,3)
	(584,0)
	(585,0)
	(586,12)
	(587,0)
	(588,0)
	(589,0)
	(590,6)
	(591,0)
	(592,0)
	(593,6)
	(594,0)
	(595,0)
	(596,7)
	(597,0)
	(598,0)
	(599,0)
	(600, 0)
}\closedcycle;
\end{axis}
\end{tikzpicture}
\caption{Distribution of operations over time in a trace.}
\label{fig:ops_per_second}
\end{figure}

Figure~\ref{fig:ops_per_second} depicts an example of the
distribution of operations in a trace file over
time~(100 nodes:~4545 pubs.,~526 subs.,~29 unpubs.,~15 unsubs.).
The rate at which operations are issued is not regular,
having occasional spikes and moments with no operations.
%
%

The area of the simulations has a rectangular shape to
mimic many of the venues we are targeting, like concert halls.
For the \gls*{dcs} version, cell size is 40x40 meters,
which entails a radio range of at least 113 meters~(roughly the
radio range of our Wi-Fi setting).
In all experiments, we use~16,~36,~64,~100,~144,~196 nodes, in
order to have an average of~2 nodes per cell, respectively, in
the following areas:~160x80m,~240x120m,~320x160m,~400x200m,
480x240m, and~560x280m.
Nodes are placed uniformly at random in the area.

Typical ad-hoc proactive routing protocols need some time for routing
tables to converge.
Thus, in our simulations, the application running on the nodes only
starts after~30 seconds.
Then, nodes begin to join the system, randomly, in the next~30 seconds.
Operations start being issued only after this.
At the end of the simulation, nodes only shutdown~60 seconds
after operations have stopped being issued.
Thus, the total simulation time is~720 seconds.
Both versions of \sys execute the same traces and use the same methodology.
%
%


\subsection{Stable and Static Nodes} 
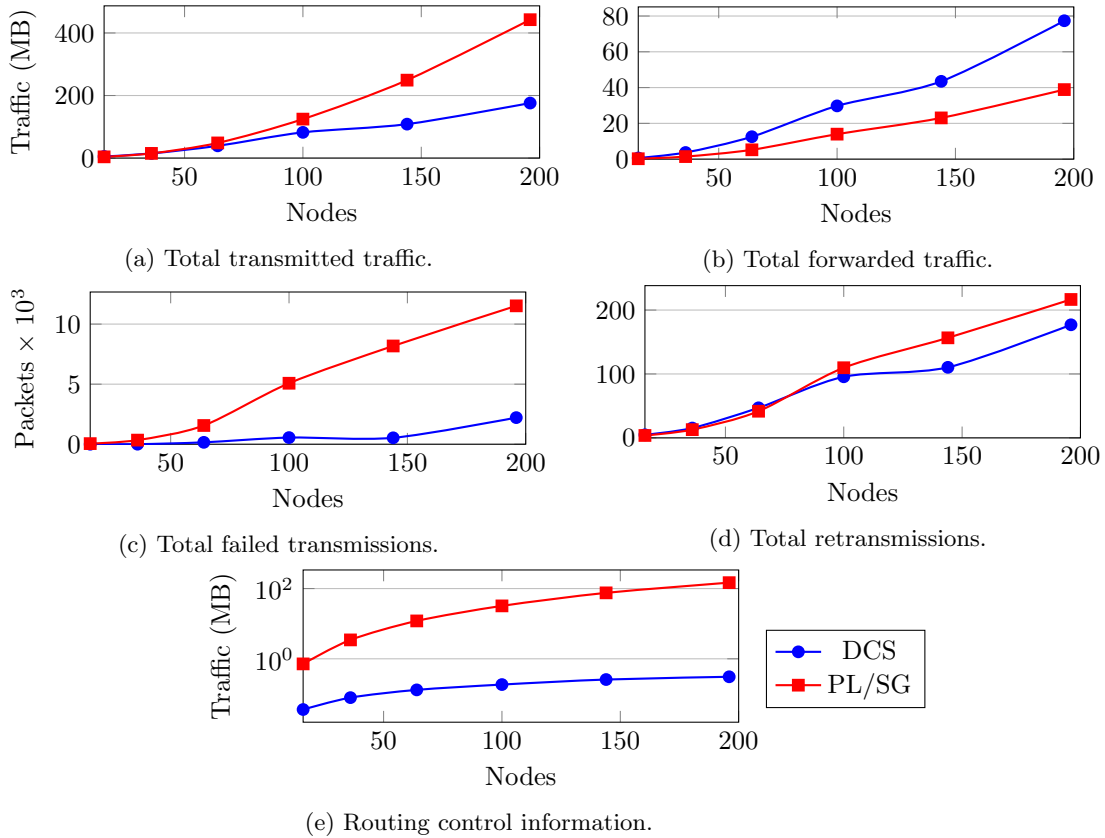
\begin{figure}[tb]
	\centering
	\begin{subfigure}{0.49\textwidth}
	\centering
	\begin{tikzpicture}
	\begin{axis}[
	xlabel=Nodes,
	ylabel=Traffic (MB),
	width=\columnwidth,
	height=3.6cm,
	xmin=16, xmax=200,
	ymin=0,
	ymajorgrids,
	legend style={at={(0.03,0.7)},anchor=west},
	legend to name=named,
	]
	\addplot[smooth, color=blue, mark=*, thick] coordinates {
		(16,	5.069531917)
		(36,	14.8142105)
		(64,	39.13717825)
		(100,	82.2202215)
		(144,	108.5080533)
		(196,	175.9286598)
	};
	\addlegendentry{DCS}
	\addplot[smooth, color=red, mark=square*, thick] coordinates {
		(16,	3.925834083)
		(36,	14.96513808)
		(64,	48.46281667)
		(100,	124.5991714)
		(144,	249.1744667)
		(196,	442.4553252)
	};
	\addlegendentry{PL/SG}
	\end{axis}
	\end{tikzpicture}
	\caption{Total transmitted traffic.}
	\label{fig:phy_tx_static}
\end{subfigure}
\hfil
\centering
\begin{subfigure}{0.49\textwidth}
	\centering
	\begin{tikzpicture}
	\begin{axis}[
	xlabel=Nodes,
	width=\columnwidth,
	height=3.6cm,
	xmin=16, xmax=200,
	ymin=0,
	ymajorgrids,
	]
	\addplot[smooth, color=blue, mark=*, thick] coordinates {
		(16,	0.6297358333)
		(36,	3.716139833)
		(64,	12.53219942)
		(100,	29.758673)
		(144,	43.49113691)
		(196,	77.36114017)
	};
	\addplot[smooth, color=red, mark=square*, thick] coordinates {
		(16,	0.22400425)
		(36,	1.431662583)
		(64,	5.218876)
		(100,	13.96954933)
		(144,	23.04903942)
		(196,	38.87371208)
	};
	\end{axis}
	\end{tikzpicture}
	\caption{Total forwarded traffic.}
	\label{fig:ip_fwd_static}
\end{subfigure}
\centering
\begin{subfigure}{0.49\textwidth}
	\centering
	\begin{tikzpicture}
	\begin{axis}[
	xlabel=Nodes,
	ylabel=Packets$\;\times\;10^3$,
	width=\columnwidth,
	height=3.6cm,
	xmin=16, xmax=200,
	ymin=0,
	ymajorgrids,
	]
	\addplot[smooth, color=blue, mark=*, thick] coordinates {
		(16,	0.00008333333333)
		(36,	0.003583333333)
		(64,	0.1580833333)
		(100,	0.5535833333)
		(144,	0.5348181818)
		(196,	2.215833333)
	};
	\addplot[smooth, color=red, mark=square*, thick] coordinates {
		(16,	0.05266666667)
		(36,	0.3440833333)
		(64,	1.559666667)
		(100,	5.075916667)
		(144,	8.1785)
		(196,	11.51883333)
	};
	\end{axis}
	\end{tikzpicture}
	\caption{Total failed transmissions.}
	\label{fig:mac_tx_failed_static}
\end{subfigure}
\hfil
\centering
\begin{subfigure}{0.49\textwidth}
	\centering
	\begin{tikzpicture}
	\begin{axis}[
	xlabel=Nodes,
	width=\columnwidth,
	height=3.6cm,
	xmin=16, xmax=200,
	ymin=0,
	ymajorgrids,
	]
	\addplot[smooth, color=blue, mark=*, thick] coordinates {
		(16,	4.7975)
		(36,	15.38275)
		(64,	47.0260)
		(100,	95.5958333333)
		(144,	110.384727273)
		(196,	176.9580)
	};
	\addplot[smooth, color=red, mark=square*, thick] coordinates {
		(16,	3.73966666667)
		(36,	12.8869166667)
		(64,	41.6844166667)
		(100,	109.742083333)
		(144,	156.5520)
		(196,	216.575583333)
	};
	\end{axis}
	\end{tikzpicture}
	\caption{Total retransmissions.}
	\label{fig:mac_tx_retrans_static}
\end{subfigure}
\centering
\begin{subfigure}{0.49\textwidth}
	\centering
	\begin{tikzpicture}
	\begin{axis}[
	xlabel=Nodes,
	ylabel=Traffic (MB),
	width=\columnwidth,
	height=3.6cm,
	xmin=16, xmax=200,
	ymin=0,
	ymode=log,
	ymajorgrids,
	scaled ticks=false, tick label style={/pgf/number format/fixed}
	]
	\addplot[smooth, color=blue, mark=*, thick] coordinates {
		(16,	0.035971)
		(36,	0.07846025)
		(64,	0.13033975)
		(100,	0.18479825)
		(144,	0.2568597273)
		(196,	0.30757)
	};
	\addplot[smooth, color=red, mark=square*, thick] coordinates {
		(16,	0.716632)
		(36,	3.454456)
		(64,	11.991478)
		(100,	32.13752)
		(144,	75.533453)
		(196,	147.427574)
	};
	\end{axis}
	\end{tikzpicture}
	\caption{Routing control information.}
	\label{fig:ip_ctl_tx_static}
\end{subfigure}
\ref*{named}
\caption{Lower layers metrics~(static).}
\label{fig:lower_layers}
\end{figure}

In Figure~\ref{fig:lower_layers}, we can observe the impact that both
materializations of \sys have on the lower layers of the network stack.
Figure~\ref{fig:phy_tx_static} reports the total traffic transmitted
by all the nodes~(at the physical layer---PHY), during the simulation.
The \gls*{plsg} version presents quite an overhead.
With~196 nodes, it reports more than the double of the
transmitted traffic by the \gls*{dcs} version.
Since we are targeting mobile devices, energy is a valuable resource.
Looking at this in an energy perspective, \gls*{plsg} will
spend twice the energy to do roughly the same work.

Figures~\ref{fig:mac_tx_failed_static} and~\ref{fig:mac_tx_retrans_static}
depict values reported by the link layer---MAC.
The Wi-Fi MAC layer implements CSMA/CA and a per hop retransmission mechanism.
In Figure~\ref{fig:mac_tx_retrans_static}, we can see the total
number of retransmitted packets.
Figure~\ref{fig:mac_tx_failed_static} shows the total number of
packets that exceeded the maximum number of retransmission attempts.
These figures show us, to some extent, the amount of interference
generated by the system itself.
The flooding strategy of \gls*{plsg}  causes a great
amount of retransmissions.
Lost messages in a wireless communication medium are inevitable.
\Gls*{dcs} also causes many retransmissions, but far
less than \gls*{plsg}.

Figure~\ref{fig:ip_fwd_static} depicts the total traffic forwarded
by every node in the system.
In some sense, this shows the amount of work nodes have to do on
behalf of the system.
In this case, \gls*{dcs} forwards more traffic because its
publish and unpublish operations also generate traffic, while
in \gls*{plsg} these are executed locally.

Figure~\ref{fig:ip_ctl_tx_static} shows the total amount
of control information the routing protocols transmit.
\Gls*{plsg} uses DSDV, a proactive routing protocol, whereas
\gls*{dcs} uses a geographic routing protocol~(a
variation of \gls*{gpsr}).
While DSDV needs to exchange lots of information to
compute the shortest paths to every other node
in the network,
%
%
the geographic routing used by \gls*{dcs}
routes messages using only local information.
Nodes only send periodic control beacons with their geographic
location~(3 bytes in size).
However, messages may be routed through longer routes.
With~196 nodes, we can see that a quarter of all the
transmitted traffic of \gls*{plsg} was control traffic.

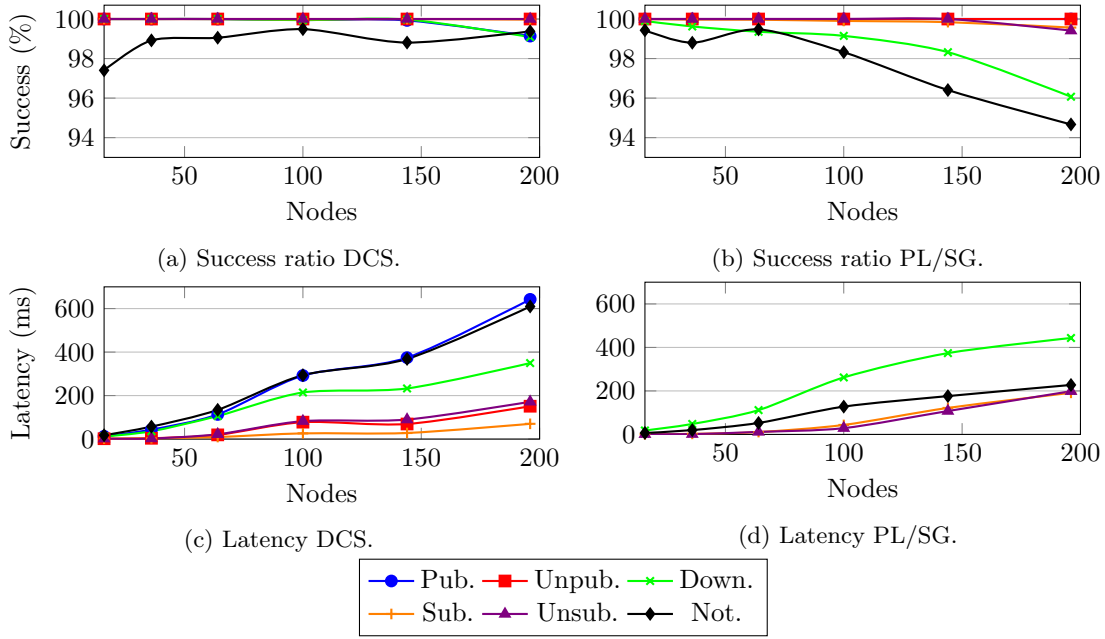
\begin{figure}[tb]
	\centering
	\begin{subfigure}{0.49\textwidth}
		\centering
		\begin{tikzpicture}
		\begin{axis}[
		xlabel=Nodes,
		ylabel=Success (\%),
		width=\columnwidth,
		height=3.6cm,
		xmin=16, xmax=200,
		ymin=93,
		ymajorgrids,
		legend columns=3,
		legend to name=named,
		]
		\addplot[smooth, color=blue, mark=*, thick] coordinates {
			(16,	100)
			(36,	100)
			(64,	100)
			(100,	99.94866153)
			(144,	99.92875463)
			(196,	99.13623021)
		};
		\addlegendentry{Pub.}
		\addplot[smooth, color=red, mark=square*, thick] coordinates {
			(16,	100)
			(36,	100)
			(64,	100)
			(100,	100)
			(144,	100)
			(196,	100)
		};
		\addlegendentry{Unpub.}
		\addplot[smooth, color=green, mark=x, thick] coordinates {
			(16,	100)
			(36,	100)
			(64,	99.99795075)
			(100,	99.94322247)
			(144,	99.97497621)
			(196,	99.08740197)
		};
		\addlegendentry{Down.}
		\addplot[smooth, color=orange, mark=+, thick] coordinates {
			(16,	100)
			(36,	100)
			(64,	100)
			(100,	99.98597082)
			(144,	100)
			(196,	99.97057958)
		};
		\addlegendentry{Sub.}
		\addplot[smooth, color=violet, mark=triangle*, thick] coordinates {
			(16,	100)
			(36,	100)
			(64,	100)
			(100,	100)
			(144,	100)
			(196,	100)
		};
		\addlegendentry{Unsub.}
		\addplot[smooth, color=black, mark=diamond*, thick] coordinates {
			(16,	97.40507203)
			(36,	98.91955839)
			(64,	99.05469547)
			(100,	99.48746532)
			(144,	98.80847536)
			(196,	99.3770346)
		};
		\addlegendentry{Not.}
		\end{axis}
		\end{tikzpicture}
		\caption{Success ratio DCS.}
		\label{fig:dcs_ops_succ_static}
	\end{subfigure}
	\hfil
	\centering
	\begin{subfigure}{0.49\textwidth}
		\centering
		\begin{tikzpicture}
		\begin{axis}[
		xlabel=Nodes,
		width=\columnwidth,
		height=3.6cm,
		xmin=16, xmax=200,
		ymin=93,
		ymajorgrids,
		]
		\addplot[smooth, color=blue, mark=*, thick, error bars/.cd,y dir=both, y explicit] coordinates {
			(16,	100) 
			(36,	100)
			(64,	100)
			(100,	100)
			(144,	100) 
			(196,	100)
		};
		\addplot[smooth, color=red, mark=square*, thick] coordinates {
			(16,	100)
			(36,	100)
			(64,	100)
			(100,	100)
			(144,	100)
			(196,	100)
		};
		\addplot[smooth, color=green, mark=x, thick] coordinates {
			(16,	99.90077951)
			(36,	99.62247461)
			(64,	99.35390249)
			(100,	99.14725674)
			(144,	98.32336429)
			(196,	96.06924808)
		};
		\addplot[smooth, color=orange, mark=+, thick] coordinates {
			(16,	100)
			(36,	99.97057131)
			(64,	99.9651255)
			(100,	99.90334008)
			(144,	99.84168244)
			(196,	99.567712)
		};
		\addplot[smooth, color=violet, mark=triangle*, thick] coordinates {
			(16,	100)
			(36,	100)
			(64,	100)
			(100,	100)
			(144,	100)
			(196,	99.41328333)
		};
		\addplot[smooth, color=black, mark=diamond*, thick] coordinates {
			(16,	99.43074116)
			(36,	98.80091315)
			(64,	99.4762098)
			(100,	98.32317234)
			(144,	96.40743219)
			(196,	94.66810782)
		};
		\end{axis}
		\end{tikzpicture}
		\caption{Success ratio PL/SG.}
		\label{fig:plsg_ops_succ_static}
	\end{subfigure}
	\centering
	\begin{subfigure}{0.49\textwidth}
		\centering
		\begin{tikzpicture}
		\begin{axis}[
		xlabel=Nodes,
		ylabel=Latency (ms),
		width=\columnwidth,
		height=3.6cm,
		xmin=16, xmax=200,
		ymin=0, ymax=700,
		ymajorgrids,
		]
		\addplot[smooth, color=blue, mark=*, thick] coordinates {
			(16,	12.77627362)
			(36,	43.50441328)
			(64,	113.2472165)
			(100,	292.1969896)
			(144,	374.400094)
			(196,	642.4375322)
		};
		\addplot[smooth, color=red, mark=square*, thick] coordinates {
			(16,	1.668685619)
			(36,	3.706974269)
			(64,	19.33324964)
			(100,	77.85985911)
			(144,	69.39011757)
			(196,	151.0790096)
		};
		\addplot[smooth, color=green, mark=x, thick] coordinates {
			(16,	11.80720982)
			(36,	35.97394641)
			(64,	105.9158378)
			(100,	214.1742375)
			(144,	232.8264584)
			(196,	349.5155771)
		};
		\addplot[smooth, color=orange, mark=+, thick] coordinates {
			(16,	1.639176106)
			(36,	3.606776358)
			(64,	9.403526715)
			(100,	26.26010017)
			(144,	27.99514204)
			(196,	70.1496244)
		};
		\addplot[smooth, color=violet, mark=triangle*, thick] coordinates {
			(16,	2.001352917)
			(36,	2.652639033)
			(64,	22.13485542)
			(100,	82.45758945)
			(144,	89.74201701)
			(196,	171.5351535)
		};
		\addplot[smooth, color=black, mark=diamond*, thick] coordinates {
			(16,	17.68751148)
			(36,	57.34119242)
			(64,	134.4235926)
			(100,	293.7731048)
			(144,	368.4026505)
			(196,	609.6963772)
		};
		\end{axis}
		\end{tikzpicture}
		\caption{Latency DCS.}
		\label{fig:dcs_ops_lat_static}
	\end{subfigure}
	\hfil
	\centering
	\begin{subfigure}{0.49\textwidth}
		\centering
		\begin{tikzpicture}
		\begin{axis}[
		xlabel=Nodes,
		width=\columnwidth,
		height=3.6cm,
		xmin=16, xmax=200,
		ymin=0, ymax=700,
		ymajorgrids,
		]
		\addplot[smooth, color=green, mark=x, thick] coordinates {
			(16,	18.11936545)
			(36,	48.07421327)
			(64,	111.8654408)
			(100,	262.0457074)
			(144,	373.9089861)
			(196,	443.3260284)
		};
		\addplot[smooth, color=orange, mark=+, thick] coordinates {
			(16,	0.5847222416)
			(36,	2.5811475)
			(64,	10.83803999)
			(100,	43.64287613)
			(144,	121.874213)
			(196,	191.1907078)
		};
		\addplot[smooth, color=violet, mark=triangle*, thick] coordinates {
			(16,	1.113489125)
			(36,	1.393288964)
			(64,	11.56980007)
			(100,	28.89937684)
			(144,	107.5465466)
			(196,	198.6705354)
		};
		\addplot[smooth, color=black, mark=diamond*, thick] coordinates {
			(16,	6.693214885)
			(36,	19.923322)
			(64,	52.80768619)
			(100,	127.825961)
			(144,	176.2701115)
			(196,	227.4556468)
		};
		\end{axis}
		\end{tikzpicture}
		\caption{Latency PL/SG.}
		\label{fig:plsg_ops_lat_static}
	\end{subfigure}
	\ref*{named}
	\caption{Application metrics~(static).}
	\label{fig:app_static}
\end{figure}

Figure~\ref{fig:app_static} depicts application level metrics.
In this static scenario, we expect neither version to surpass the other.
Regarding operation success
ratio~(Figures~\ref{fig:dcs_ops_succ_static}
and~\ref{fig:plsg_ops_succ_static}), we verify that
both versions have more than~94\% success for every operation.
In both versions, notifications are a type of message
that does not employ an application level retransmission
mechanism~(\S\ref{ssec:retrans}).
Thus, they are more susceptible to interferences, and can
end up being lost.
%

Regarding operation latency~(Figures~\ref{fig:dcs_ops_lat_static}
and~\ref{fig:plsg_ops_lat_static}), we can see the advantage
of calculating shortest paths.
Notification operations have lower latency in \gls*{plsg},
because the geographic routing of \gls*{dcs} cannot compete with
the shortest paths of DSDV\@.
On the other hand, download operations in \gls*{dcs} have a
slightly lower latency, because \gls*{dcs} causes overall less
interferences and it employs a location-aware strategy when
retrieving data~(\S\ref{ssec:down2}).
%
%



\subsection{Static but Failing Nodes} 
Regarding churn, \ie, the ingress and egress of nodes in
the system, we experiment with two different scenarios.
We show the impact of nodes leaving the system
definitely, \eg, nodes crashing.
Secondly, we show the impact of nodes with intermittent failures
thus, entering and leaving the system multiple times over time.
These scenarios allow to evaluate aspects regarding data
availability and persistence in the presence of node failures.

\subsubsection{Permanent Failures}
In this scenario, nodes are either publishers or subscribers, and
publishers choose a random instant~(between~200 and~300 seconds of
the simulation) to leave the system abruptly.

\begin{figure}[tb]
	\centering
	\begin{tikzpicture}
	\begin{axis}[
	xlabel=Failing Nodes (\%),
	ylabel=Success (\%),
	width=0.8\textwidth,
	height=3.6cm,
	xmin=5, xmax=50,
	legend style={at={(0.7,0.55)},anchor=west},
	ymajorgrids,
	]
	\addplot[smooth, color=blue, mark=*, thick] coordinates {
		(5,	97.72713627)
		(10,	97.52024267)
		(20,	98.40956881)
		(30,	98.18061843)
		(40,	97.8732327)
		(50,	97.02108594)
	};
	\addlegendentry{DCS}
	\addplot[smooth, color=red, mark=square*, thick] coordinates {
		(5,	98.08542085)
		(10,	96.52265088)
		(20,	92.82077061)
		(30,	88.43015471)
		(40,	83.11543246)
		(50,	81.76068852)
	};
	\addlegendentry{PL/SG}
	\end{axis}
	\end{tikzpicture}
	\vspace{-7pt}
	\caption{Notifications success ratio (crashing, 100 nodes).}
	\label{fig:nots_crash}
\end{figure}
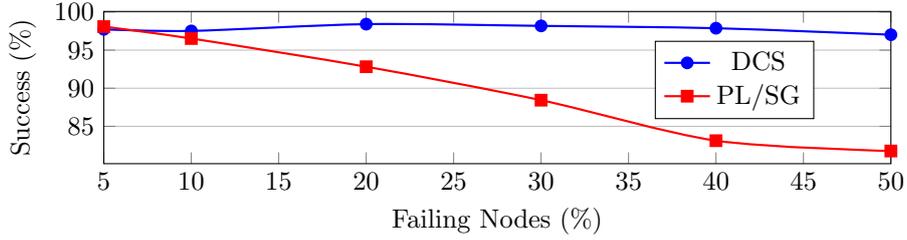

In \gls*{plsg}, publications are executed locally.
This can be an advantage, since it does not require communication.
But it can also be a disadvantage, since only the publisher keeps that data.
If that node fails, all the data it stored will disappear with it.
Figure~\ref{fig:nots_crash} shows exactly that.
As the amount of failing nodes increases, in \gls*{plsg},
the nodes with the published data leave the system thus, the
matching between subscriptions and publications is not detected.
However, since \gls*{dcs} employs replication
mechanisms~(\S\ref{ssec:rep}), even when the publisher
nodes leave the system, the matching still occurs.

\subsubsection{Transient Failures}
In this scenario, nodes are selected at random.
The selected nodes alternate between periods in the on and off states.
Nodes are on for periods of~120 seconds, and are off for
periods of~60 seconds.
When changing state, nodes have a probability of~75\% of
changing to the opposite state, otherwise they stay in the same state.

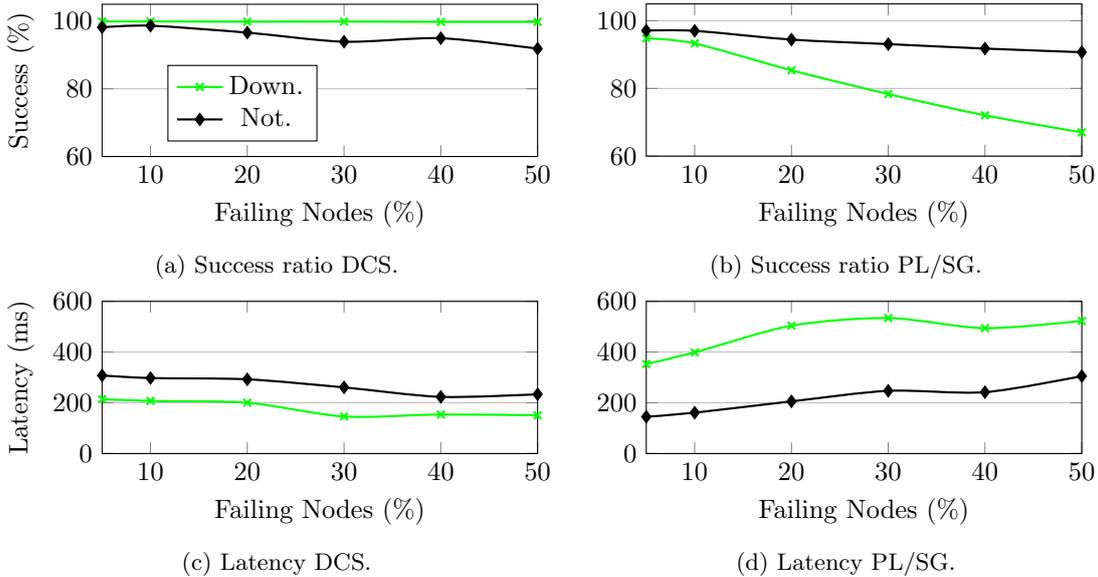
\begin{figure}[tb]
	\centering
	\begin{subfigure}{0.49\textwidth}
		\centering
		\begin{tikzpicture}
		\begin{axis}[
		xlabel=Failing Nodes (\%),
		ylabel=Success (\%),
		width=\columnwidth,
		height=3.6cm,
		xmin=5, xmax=50,
		ymin=60, ymax=105,
		ymajorgrids,
		legend style={at={(0.15,0.35)},anchor=west},
		legend columns=1,
		]
		\addplot[smooth, color=green, mark=x, thick] coordinates {
			(5,	99.90973154)
			(10,	99.91781132)
			(20,	99.83446155)
			(30,	99.87329575)
			(40,	99.79237955)
			(50,	99.79135768)
		};
		\addlegendentry{Down.}
		\addplot[smooth, color=black, mark=diamond*, thick] coordinates {
			(5,	98.19613335)
			(10,	98.66548689)
			(20,	96.61361053)
			(30,	93.92733566)
			(40,	94.96361968)
			(50,	91.86021444)
		};
		\addlegendentry{Not.}
		\end{axis}
		\end{tikzpicture}
		\caption{Success ratio DCS.}
		\label{fig:dcs_ops_succ_inter}
	\end{subfigure}
	\hfil
	\centering
	\begin{subfigure}{0.49\textwidth}
		\centering
		\begin{tikzpicture}
		\begin{axis}[
		xlabel=Failing Nodes (\%),
		width=\columnwidth,
		height=3.6cm,
		xmin=5, xmax=50,
		ymin=60,ymax=105,
		ymajorgrids,
		]
		\addplot[smooth, color=green, mark=x, thick] coordinates {
			(5,	94.85164554)
			(10,	93.31154967)
			(20,	85.38059381)
			(30,	78.34216003)
			(40,	72.0581711)
			(50,	67.01631218)
		};
		\addplot[smooth, color=black, mark=diamond*, thick] coordinates {
			(5,	97.11653884)
			(10,	97.0292864)
			(20,	94.44588439)
			(30,	93.10973744)
			(40,	91.79289352)
			(50,	90.72024795)
		};
		\end{axis}
		\end{tikzpicture}
		\caption{Success ratio PL/SG.}
		\label{fig:plsg_ops_succ_inter}
	\end{subfigure}
	\centering
	\begin{subfigure}{0.49\textwidth}
		\centering
		\begin{tikzpicture}
		\begin{axis}[
		xlabel=Failing Nodes (\%),
		ylabel=Latency (ms),
		width=\columnwidth,
		height=3.6cm,
		xmin=5, xmax=50,
		ymin=0, ymax=600,
		ymajorgrids,
		legend columns=3,
		]
		\addplot[smooth, color=green, mark=x, thick] coordinates {
			(5,	214.4258561)
			(10,	207.3402621)
			(20,	200.3707645)
			(30,	146.0055147)
			(40,	153.9330626)
			(50,	150.6477422)
		};
		\addplot[smooth, color=black, mark=diamond*, thick] coordinates {
			(5,	307.7803689)
			(10,	297.5204835)
			(20,	292.5898794)
			(30,	260.7189912)
			(40,	223.4008424)
			(50,	233.7124891)
		};
		\end{axis}
		\end{tikzpicture}
		\caption{Latency DCS.}
		\label{fig:dcs_ops_lat_inter}
	\end{subfigure}
	\hfil
	\centering
	\begin{subfigure}{0.49\textwidth}
		\centering
		\begin{tikzpicture}
		\begin{axis}[
		xlabel=Failing Nodes (\%),
		width=\columnwidth,
		height=3.6cm,
		xmin=5, xmax=50,
		ymin=0, ymax=600,
		ymajorgrids,
		]
		\addplot[smooth, color=green, mark=x, thick] coordinates {
			(5,	353.0412557)
			(10,	398.7417192)
			(20,	503.9250279)
			(30,	534.0218701)
			(40,	494.1039899)
			(50,	522.7266511)
		};
		\addplot[smooth, color=black, mark=diamond*, thick] coordinates {
			(5,	145.0571141)
			(10,	161.1459638)
			(20,	205.7126287)
			(30,	247.2593513)
			(40,	241.9350126)
			(50,	305.2377222)
		};
		\end{axis}
		\end{tikzpicture}
		\caption{Latency PL/SG.}
		\label{fig:plsg_ops_lat_inter}
	\end{subfigure}
	\caption{Application metrics~(transient, 100 nodes).}
	\label{fig:app_inter}
\end{figure}

With nodes entering and leaving the system constantly, the operations that can be more affected are the notifications detection and the downloads.
Figure~\ref{fig:app_inter} presents the application metrics for these two operations.
Once again, since \gls*{dcs} employs replication mechanism, it is little affected by the intermittent churn~(Figures~\ref{fig:dcs_ops_succ_inter} and~\ref{fig:dcs_ops_lat_inter}).
However, \gls*{plsg} suffers from low success rate in the download operations~(Figures~\ref{fig:plsg_ops_succ_inter}).
Although the notifications are detected, when a node tries to download the data, as the amount of failing nodes increases, the probability of the data owner being off also increases.
This is also accompanied by an increase in the latency of notifications~(Figure~\ref{fig:plsg_ops_lat_inter}).
In \gls*{plsg}, the matching between a node's publications and subscriptions that were issued when the node was off have to wait for the node to switch state and join the system~(\S\ref{sec:plsg}).
When joining the system, in \gls*{plsg}, a node receives the subscriptions issued by all the other nodes previous to it entering the system.
Then, the joining node finds the new subscriptions it received and checks if it has matching publications.

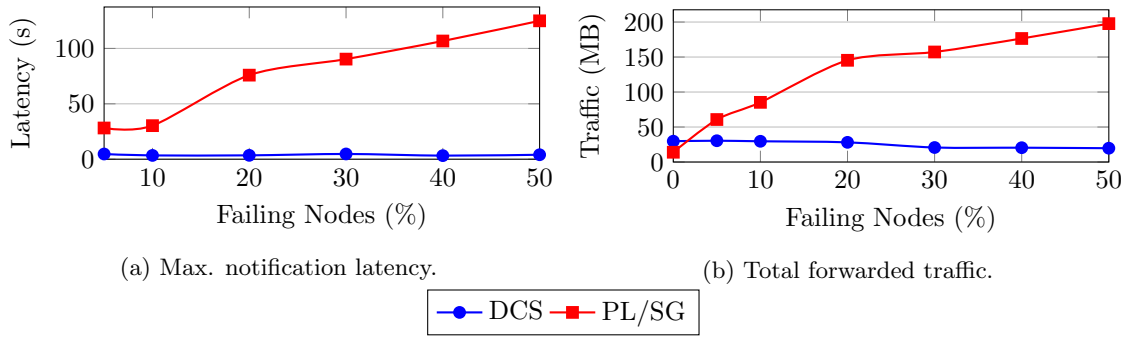
\begin{figure}[tb]
	\centering
	\begin{subfigure}{.49\textwidth}
	\begin{tikzpicture}
	\begin{axis}[
	xlabel=Failing Nodes (\%),
	ylabel=Latency (s),
	width=\columnwidth,
	height=3.6cm,
	xmin=5, xmax=50,
	ymin=0,
	legend columns=2,
	legend to name=named,
	ymajorgrids,
	]
	\addplot[smooth, color=blue, mark=*, thick] coordinates {
		(5,	4.692860401)
		(10,	3.413390531)
		(20,	3.487964429)
		(30,	4.732818809)
		(40,	3.224197199)
		(50,	4.029479343)
	};
	\addlegendentry{DCS}
	\addplot[smooth, color=red, mark=square*, thick] coordinates {
		(5,	28.17953865)
		(10,	30.29417616)
		(20,	75.89048859)
		(30,	90.35850281)
		(40,	106.7043727)
		(50,	124.9221708)
	};
	\addlegendentry{PL/SG}
	\end{axis}
	\end{tikzpicture}
	\caption{Max. notification latency.}
	\label{fig:nots_max_ops_lat_inter}
\end{subfigure}
\hfil
	\centering
	\begin{subfigure}{.49\textwidth}
	\begin{tikzpicture}
	\begin{axis}[
	xlabel=Failing Nodes (\%),
	ylabel=Traffic (MB),
	width=\columnwidth,
	height=3.6cm,
	xmin=0, xmax=50,
	ymin=0,
	ymajorgrids,
	]
	\addplot[smooth, color=blue, mark=*, thick] coordinates {
		(0,	29.758673)
		(5,	30.41239)
		(10,	29.63843678)
		(20,	28.11600538)
		(30,	20.73560756)
		(40,	20.46327)
		(50,	19.73661778)
	};
	\addplot[smooth, color=red, mark=square*, thick] coordinates {
		(0, 13.96954933)
		(5,	60.71779867)
		(10,	85.26647725)
		(20,	145.3496318)
		(30,	157.2171159)
		(40,	176.4942288)
		(50,	197.7343028)
	};
	\end{axis}
	\end{tikzpicture}
	\caption{Total forwarded traffic.}
	\label{fig:ip_fwd_inter}
\end{subfigure}
\centering
\ref*{named}
\caption{Transient scenario, 100 nodes.}
\label{fig:inter}
\end{figure}

Figure~\ref{fig:nots_max_ops_lat_inter} corroborates this, even further.
The maximum latency for \gls*{dcs} notifications stays stable
as the amount of failing nodes increases.
But, in \gls*{plsg}, the maximum latency for a notification
increases to values around 100 seconds with~40\% of failing nodes.

Figure~\ref{fig:ip_fwd_inter} shows a byproduct of
the downloads low success ratio.
With no churn, \gls*{dcs} forwards more traffic because
its publish and unpublish operations require communication.
However, with this kind of intermittent churn, \gls*{plsg}
forwards much more traffic than \gls*{dcs}.
This is due to the fact that download operations are
retried~(and fail) many times.
This entering and leaving of nodes from the network
causes routing tables to become out of date, and thus
need to change more frequently.

\subsection{Stable but Mobile Nodes} 
When moving, nodes use the \gls*{rwp} mobility model, which interleaves
pauses with movement.
We argue that the plain \gls*{rwp} mobility model does
not quite mimic the movement pattern people have in the kind
of events we target.
For instance, in a football game or a music concert, people do not
move constantly.
In fact, they do not move much during the majority of the
event, except during intermissions.
Thus, to make it more better resemble our target scenarios, we
made an adaptation: every time a node is about to move, it
tosses a coin do decide whether to move or not.
%
If not moving, the node continues once again in the pause moment.

For this scenario, we consider three maximum speeds:~0.6~m/s, 1.4~m/s,
and~2.5~m/s~(slow walking, regular walking, and running, respectively).
Only~60\% of nodes are mobile, and have a moving probability of~80\%.

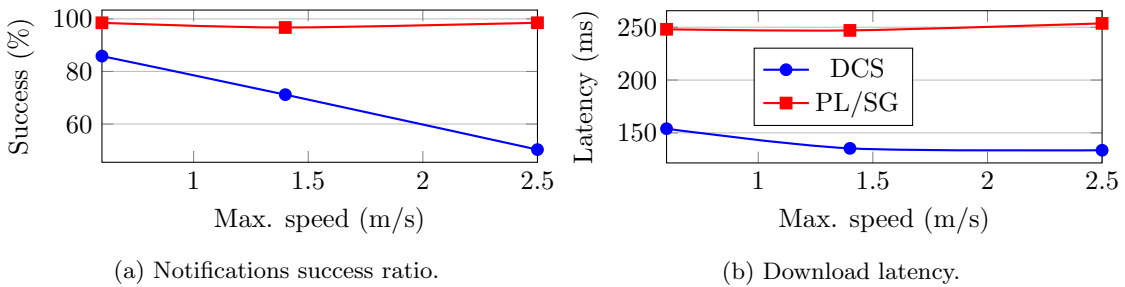
\begin{figure}[tb]
\begin{subfigure}{0.49\textwidth}
	\centering
	\begin{tikzpicture}
	\begin{axis}[
	xlabel=Max. speed (m/s),
	ylabel=Success (\%),
	width=\columnwidth,
	height=3.6cm,
	xmin=0.6, xmax=2.5,
	ymajorgrids,
	]
	\addplot[smooth, color=blue, mark=*, thick] coordinates {
		(0.6,	85.83910048)
		(1.4,	71.17841266)
		(2.5,	50.27315297)
	};
	\addplot[smooth, color=red, mark=square*, thick] coordinates {
		(0.6,	98.51554244)
		(1.4,	96.72325714)
		(2.5,	98.5325327)
	};
	\end{axis}
	\end{tikzpicture}
	\caption{Notifications success ratio.}
	\label{fig:ops_succ_mob}
\end{subfigure}
\centering
\begin{subfigure}{0.49\textwidth}
	\centering
	\begin{tikzpicture}
	\begin{axis}[
	xlabel=Max. speed (m/s),
	ylabel=Latency (ms),
	width=\columnwidth,
	height=3.6cm,
	xmin=0.6, xmax=2.5,
	legend style={at={(0.2,0.5)},anchor=west},
	ymajorgrids,
	]
	\addplot[smooth, color=blue, mark=*, thick] coordinates {
		(0.6,	153.9681228)
		(1.4,	135.3472375)
		(2.5,	133.5636548)
	};
	\addlegendentry{DCS}
	\addplot[smooth, color=red, mark=square*, thick] coordinates {
		(0.6,	248.0442361)
		(1.4,	246.9698171)
		(2.5,	253.6225514)
	};
	\addlegendentry{PL/SG}
	\end{axis}
	\end{tikzpicture}
	\caption{Download latency.}
	\label{fig:ops_lat_mob}
\end{subfigure}
\vspace{-7pt}
\caption{Mobile scenario, 100 nodes, pause 120 seconds.}
\label{fig:mob}
\end{figure}

Figure~\ref{fig:ops_succ_mob} shows some caveats of \gls*{dcs}.
We can see that increasing the speed lowers the notifications success ratio.
We claim this happens because 
every node inside a cell is supposed to have the same state and work collaboratively as one.
But, the intra-cell communication primitive is the one-hop broadcast, that
is unreliable by nature.
Thus, nodes inside a cell may not receive the same messages.
This makes that, nodes inside the same cell can have different cell states, and thus reply different answers to the same ``question''.
Mobility creates even more entropy in this cell state.

Figure~\ref{fig:ops_lat_mob} presents a byproduct of the location-aware download strategy used by \gls*{dcs}.
While, \gls*{plsg} is required to download data from their owners, \gls*{dcs} might have different replicas for download at its disposal, and it can choose the one closer to the requester.

\subsection{Discussion}
Due to its flooding approach, \gls*{plsg} causes far more interferences than \gls*{dcs}.
This is exacerbated the larger is the network size.
Churn is also a problem for \gls*{plsg}, because publications are executed locally.
In summary, \gls*{plsg} is more suitable for smaller scenarios with no data availability requirements.

In turn, \gls*{dcs} takes advantage of its geographic routing to employ replication and a location-aware data retrieval strategy.
However, since one-hop broadcast is unreliable by nature, the assumption that every node inside a cell have the same state, needs to be relaxed.
Thus, \gls*{dcs} is more suitable for larger scenarios with low mobility.
%
%





\glsresetall
\section{Related Work}
\label{sec:relwork}
Much work has already been done in \acrshort*{ps} systems, both for
wired and wireless settings.
However, the notions of time or persistence have not been addressed in most.
In wired environments, some \acrshort*{ps} systems 
 explore the notion of a
persistent data repository to
support the delivery of notifications to disconnected
clients.
For this, they use proxy servers  that
maintain permanent connections to
the broker network, and buffer any notifications received while
clients are disconnected~\cite{Sutton:2001}.
The main drawback is that clients need to reconnect
to the same proxies to receive the buffered notifications.

In~\cite{Cilia:2003}, the authors propose a  \acrshort*{ps}
system  that allows the retrieval of data from the past.
Clients must indicate how many data items from the past
they want to get.
This data is replicated across several
\textit{buffer} nodes
and may have to
be collected from many of these.

In wireless settings, 
Chapar~\cite{Khakpour:2010} is a
\acrshort*{ps} system for mobile ad-hoc networks.
It uses a  broker network (based on an OLSR overlay) to handle publications
and subscriptions.
It also allows notifications to be buffered in replicated
data containers until their expiration time elapses or
they are delivered to all their intended subscribers.

GeoRendezvous~\cite{Carvalho:2006} is a \acrshort*{ps} system for
wireless networks that also makes use of a
cluster-based \acrshort*{ght}. 
%
However, it does not provide time-awareness nor publication persistence.
%
It uses multiple hash
functions to hash topics to different cells and allow
subscribers to choose the closest cells to themselves.
This is very similar to the way \sys uses the \acrshort*{ght} to
download objects from the closest replicas. 
Contrary to \sys, it only allows one tag per subscription.
%
%

%
%

Regarding other   data dissemination mechanisms for wireless networks,
Krowd~\cite{Drolia:2015} provides a key/value store 
 for sharing data among nearby
wireless devices.
It deploys a one-hop 
\acrshort*{dht}, where each device is
connected to every other device,  
requiring complete knowledge of the network.
%
%
It does not provide  data replication, and is not
resistant to device mobility or churn.
It also does not address data availability in face of device failure.

Ephesus~\cite{silva2016ephemeral} is a decentralized
 key/value store  for networks of mobile
devices.
%
%
It builds on a non-mobility-aware \acrshort*{dht}
to tolerate churn, uses data replication to address data
persistence and availability,  and does not require   knowledge on the entire network.

 iTrust~\cite{michel2013mobile}  is
a distributed system to publish and retrieve content
from mobile ad-hoc networks.
%
To publish content, a device sends the item's
 whereabouts  to a
random set of remote peers.
To obtain a content, it contacts
 a random set of  neighbors until it finds one that knows the content's location.
%
%
Data availability is not addressed.
%
%

PDS~\cite{7979975} is the work that more closely relates to ours.
However, it adopts a query/response interaction model,
rather than   \acrshort*{ps}.
PDS is designed for content-centric data discovery and
retrieval on opportunistically gathered  edge devices.
It targets small scale networks with low to moderate mobility.
%
Data can be widely
cached at any willing and capable device, which 
 can lead to serious storage overheads.
Another important difference is that PDS does not have the
notion of ``publishing content''.
Data is only distributed/cached if some device requests it.
As a result
only popular  items have some availability guarantees, while less popular data may even disappear.
Moreover, users must proactively  search for content they
want. 
%
In \sys, users ``install'' their  subscriptions, and as long
as these are active, users will keep receiving
notifications of  the content published by others.
%
%



\section{Conclusions}
\label{sec:conc}
In this paper, we present the concept of a \gls*{tps}
system where subscriptions have an associated time
scope that can reside either in the future, present or past.
Thus, requiring both subscriptions and publications
to be persistent in the system.
In the end, this conceptually merges a \gls*{ps} system
with a storage system.
We also present the design of \sys, a novel extended
topic-based \gls*{tps} system for wireless networks of
mobile devices.
We present two different materializations of \sys: a
simplistic approach using a \gls*{plsg} rationale;
and a more intricate one, following a \gls*{dcs} approach
using a \gls*{ght} as a storage substrate.
This work can be seen has a first step towards a
data dissemination/sharing system for a wide-area
setting like a campus, or a football stadium.

As future work,
we highlight a
proof of concept implementation of \sys for networks of Android
mobile devices~\cite{cerqueira:mecc:2017}, and the
integration of this approach with infrastructure support
whenever possible~\cite{silva:mecc:2016}.
As future directions we highlight privacy and security
concerns in this type of networks~(mainly access control), and
tackling the problem of handling large data items.

\section*{Acknowledgement}
This work was partially supported by FCT-MCTES, through
project Hyrax~(CMUP-ERI/FIA/0048/2013), NOVA LINCS research
center~PEst-UID/CEC/04516/2013, and grant SFRH/BD/99486/2014; and
by the European Union, through project LightKone~(grant agreement nº732505).

\bibliographystyle{plain}
\bibliography{bibliography}

\end{document}